\documentstyle[floats,aps]{revtex}
\begin{document}
\title{Quantum gravity corrections to neutrino propagation}
\author{
Jorge Alfaro$^a$\thanks{jalfaro@chopin.fis.puc.cl},
Hugo A. Morales-T\'ecotl$^b$\thanks{Associate member of Abdus-Salam
ICTP, Trieste, Italy. hugo@xanum.uam.mx}
and Luis F. Urrutia$^c$\thanks{me@aurora.nuclecu.unam.mx}
}
\address{
$^a$Facultad de F\'{\i}sica,  Pontificia Universidad Cat\'olica de  
Chile, Casilla 306, Santiago 22, Chile\\
$^b$Departamento de F\'{\i}sica,  Universidad Aut\'onoma Metropolitana
Iztapalapa \\  A.P. 55-534, M\'exico D.F. 09340, M\'exico\\
$^c$Departamento de F\'\i sica de Altas Energ\'\i as, Instituto de Ciencias Nucleares \\    Universidad Nacional Aut\'onoma
de M\'exico,  A.P. 70-543, M\'exico D.F. 04510, M\'exico
}
\date{September 24 1999}
\maketitle
\begin{abstract}
Massive spin-1/2 fields are studied in the framework of loop quantum gravity by considering a state approximating, at a length scale $\cal L$ much greater than Planck length $\ell_P=1.2\times 10^{-33}$cm, a spin-1/2 field in flat spacetime. The discrete structure of spacetime at $\ell_P$ yields corrections to the field propagation at scale $\cal L$. Next, Neutrino Bursts (${\bar p}\approx 10^5$GeV) accompaning Gamma Ray Bursts that have travelled cosmological distances, $L\approx 10^{10}$ l.y., are considered.
The dominant correction is helicity independent and leads to a time delay w.r.t. the speed of light, $c$, of order $({\bar p} \ \ell_P) L/c\approx 10^4$s. To next order in ${\bar p} \ \ell_P$ the correction has the form of the Gambini and Pullin effect for photons.   
Its contribution to time delay is comparable to that caused by the mass term. Finally, a dependence $L_{\rm os}^{-1} \propto {\bar p}^2 \ell_P$ is found for a two-flavour neutrino oscillation length.

\end{abstract}

The fact that  some Gamma Ray Bursts (GRB) originate at cosmological distances, ($\approx 10^{10} \ {\rm light \ years})$ \cite{METZ}, together with time resolutions down to submillisecond scale  achieved in recent GRB data \cite{BHAT}, suggest that it is possible to probe fundamental laws of physics at energy scales near to Planck energy $E_P=1.2\times 10^{19}$ GeV \cite{AC,ELLIS}. Furthermore, sensitivity will be improved with HEGRA and Whipple air Cerenkov telescopes and by AMS and GLAST spatial experiments. Thus, quantum gravity effects could be at the edge of observability \cite{AC,ELLIS}. 
Now, quantum gravity theories imply different spacetime structures \cite{ELLIS,RROV} and it can be expected that what we consider flat spacetime, can actually involve dispersive effects  arising from Planck scale lengths. Such tiny effects might become observable upon accumulation over travels through cosmological distances by energetically enough particles like cosmological GRB photons.  

Now, the most widely accepted model of GRB,
so called fireball model, predicts also  the generation of $10^{14}-10^{19}$ eV Neutrino Bursts (NB) \cite{WAX,VIETRI}. Yet, another GRB model based on cosmic strings requires neutrino production \cite{PLA}. 
Present experiments  to observe high energy astrophysical neutrinos like AMANDA, NESTOR, Baikal, ANTARES and Superkamiokande, for example, will detect at best only one or two neutrinos
in coincidence with GRB's per year. The planned  Neutrino Burster Experiment (NuBE) will measure the flux of ultra high energy neutrinos ($>$ 10 TeV)  over a $\sim 1{\rm km}^2$ effective area, in coincidence with satellite measured GRB's \cite{ROY}. It is expected to detect $\approx$ 20 events per year, according to the fireball model. Hence, one can study quantum gravity effects on astrophysical neutrinos that might be observed or, the other way around, such observations could  be used to restrict quantum gravity theories.

In this letter, the loop quantum gravity framework is adopted. In this context, Gambini an Pullin  studied light propagation semiclassically \cite{GP}. They found, besides departures from perfect non-dispersiveness of ordinary vacuum, helicity depending corrections for propagating waves. In the present work, the case of massive spin-$1/2$ particles in loop quantum gravity is studied also semiclassically. They could be identified with the neutrinos that would be produced in GRB. Central ideas  and results are presented whereas details will appear elsewhere \cite{AMU}.

Loop quantum gravity \cite{RROV} uses a spin networks basis, labelled by graphs embedded in a three dimensional space $\Sigma$. Physical predictions hereby obtained are a ``polymer-like"  structure of space \cite{geooperator} and a possible explanation of black hole entropy \cite{Sbh}.  A first attempt to couple spin-$\frac{1}{2}$ fields to gravity, along these lines, was made in \cite{HMROV} and a generalization to the spin networks basis has been developed in \cite{fsnet}.
A significant progress in  the loop approach to quantum gravity  was made by Thiemann, who put forward a consistent regularization procedure to properly define the  quantum Hamiltonian constraint of the full theory, which includes the Einstein plus matter (leptons, quarks, Higgs particles) contributions \cite{Thiemann}. It is based on a triangulation of space with tetrahedra whose sides  are of the order $\ell_P$. The cornerstone of Thiemann's proposal is the incorporation of the volume operator as a convenient regulator, since its action upon states is finite. Having at our disposal a regularized version of the quantum Hamiltonian describing fermions coupled to Einstein gravity, we will further need a loop state which approximates a flat 3-metric on $\Sigma$, at scales ${\cal L}$ much larger than  the Planck length. For pure gravity this state is called weave \cite{weave}. A flat weave $|W\rangle$ is characterized by a length  scale ${\cal L} > \!\! > \ell_P$, such that for distances $ d \geq {\cal L}$ the continuous flat classical geometry is regained, while for  distances $ d <\!\!< {\cal L} $ the  quantum loop structure of space is manifest.
The stronger the unequality $ d>\!\!>\ell_P$ 
holds, the more isotropic and homogeneous the weave looks like. For example, the metric operator ${\hat g}_{ab}$ satisfies $\langle W | {\hat g}_{ab}| W\rangle  = \delta_{ab} + {\cal O} (\frac{\ell_P}{{\cal L}})$. Now, a generalization of such an idea to include matter fields is required. For our analysis it suffices to exploit the main features that  a flat weave with fermions must have: in particular, it must reproduce Dirac equation in flat spacetime and this is just the basis of our approximation scheme. It is denoted by $|W, \xi\rangle$, has a characteristic length ${\cal L}$ and it is referred to simply as a weave.

The use of Thiemann's regularization for Einstein Dirac theory 
naturally allows the semiclassical treatment here pursued; expectation values w.r.t. $|W,\xi\rangle$ are considered thereby. They are expanded around relevant vertices of the triangulation and a systematic approximation is given involving the scales $\ell_P <\!\!< \lambda_D <\!\!<  \lambda_C$,
the last two corresponding to, respectively,  
De Broglie  and Compton wavelengths of a light fermion. Corrections come out at this level.


The Hamiltonian constraint for a spin-$\frac{1}{2}$ field coupled  
to gravity consists of  a pure gravity contribution,
a kinetic fermion term, namely,
\begin{eqnarray}
{H}^{(1)} &:=& \int d^3x   \frac{E_i{}^a}{2\sqrt{det(g)}}
\left(i\pi^T\tau_i{\cal D}_a\xi + c.c. \right)\label{h12} \;, 
\label{H1}
\end{eqnarray}
and other terms \cite{Thiemann} whose contribution is summarized in (\ref{effnueq}) below. Their analysis is an extension of the one for $H^{(1)}$
given here and will be spelled out in \cite{AMU}.
We use ${\vec \tau}= -\frac{i}{2} {\vec \sigma}$, 
the latter being the standard Pauli matrices. 
The fermion field is a Grassmann valued Majorana spinor
$\Psi^T= (\psi^T, (-i{\sigma^2} {\psi^*})^T)$.
The two component spinor $\psi$ has definite chirality and 
it is a scalar under general coordinate transformations. Hence, (\ref{H1}) is not parity invariant. The configuration variable is $\xi= (det(q))^\frac{1}{4} \psi$, which is a half density. The corresponding momentum is, with this choice, $\pi = i \xi^*$, similarly as in flat space. The gravitational canonical pair consists of $E_i{}^a$ and the $su(2)$-connection $A^j_b$ (${\cal D}$), where
$(detg)g^{ab}= {E_i{}^a}E^{ib}$.

Upon regularization \cite{Thiemann}, the expectation value of (\ref{H1}) with respect to the weave becomes 
\begin{eqnarray}
\label{EXPW}
\langle W,\xi |\hat{H}^{(1)}|W,\xi \rangle &=& 
-\frac{\hbar}{4\ell_P^4} \sum_{v\in V(\gamma)}\frac{8}{E(v)}
\epsilon^{ijk}\sum_{s_I\cap s_J\cap s_K=v}\epsilon^{IJK} \times \nonumber\\
&& \mbox{}
\left\{ 
\langle W,\xi|\hat\xi_B(v + s_K(\Delta))\frac{\partial\;}{\partial\xi^A(v)} 
(\tau_k h_{s_K(\Delta)})^{AB} \hat{w}_{iI\Delta}(v) \hat{w}_{jJ\Delta}(v)
|W,\xi \rangle \right. \nonumber\\
&& \mbox{}\left. - \langle W,\xi|\tau_k\hat\xi(v)\frac{\partial\;}{\partial\xi(v)}| 
 \hat{w}_{iI\Delta}(v) \hat{w}_{jJ\Delta}(v)|W,\xi \rangle
-c.c. \right\}.
\end{eqnarray}
Here an adapted triangulation of $\Sigma$ to the graph $\gamma$ of the weave state $|W,\xi \rangle$ is adopted. Auxiliary quantities used are
${\hat w}_{k I \Delta}= Tr\left( \tau_k h_{s_I(\Delta)}\left[ h^{-1}_{s_I(\Delta)},\sqrt{V_v}\right] \right)$, where $V_v$ is the volume operator restricted to act upon vertex $v$. 
$h_{s(\Delta)}$ are holonomies along segments, $s$, of edges forming tetrahedra in the triangulation $\Delta$ \cite{Thiemann}.
$V(\gamma)$ stands for the set of vertices of $\gamma$. The second sum,
$\sum_{s_I\cap s_J\cap s_K=v}$, involves triples of segments
$s_I, s_J, s_K$ intersecting at $v$.
Notice that one actually averages over 
$E(v)= n_v(n_v-1)(n_v-2)/6$ possible triangulations (one for each triple of edges) when the vertex $v$ is reached by $n_v$  edges (the valence) of the graph.  

To estimate (\ref{EXPW}) we associate  to it  c-number quantities  respecting the index structure,  together with appropriate scale factors arising  from dimensional reasons and, most important, in line with the weave state approximating flat space with fermions. This amounts to an {\em expansion of expectation values} around vertices of the weave. The explicit form is taken from the expansion  the involved operators would have in powers of the segments $s^a, |s^a|\sim\ell_P$
--a procedure justified for weave states.
Useful quantities coming  in by expanding   
$\hat{w}_{iI\Delta}(v)= s_I{}^a\hat{w}_{ia}+ s_I{}^as_I{}^b\hat{w}_{iab} +\dots$, for instance, are
\begin{equation}
\label{DEFW}
\hat w_{ia}= \frac{1}{2} [{A}_{ia}, \sqrt{V_v}], \qquad
\hat w_{i a b}=\frac{1}{8} \epsilon_{ikl} \  [ {A}_{k a}, [ A_{l b}, \sqrt{V_v}]]\;,
\end{equation}
whose contribution to the average in the weave is estimated by considering that of ${A}_{ia}$ and $\sqrt{V_v}$ to be of order $\sim 1/{\cal L}$ and $\sim \ell_P^{3/2}$, respectively.
To proceed with the approximation we think of space as made up of boxes of volume ${\cal L}^3$, whose center is denoted by ${\vec x}$. Each box contains a large number of 
vertices of the weave, but is considered infinitesimal in the scale where the space can be regarded as continuous, so that we take ${\cal L}^3\approx d^3x$. Let ${\hat F}(v)$ be a fermionic operator which produces the slowly varying (inside the box) funtion $F({\vec x})$; i.e. ${\cal L}<\!\!<\lambda_D$. Also let $\frac{1}{\ell_P^3}\hat{G}(v)$ be a gravitational operator with average within the box $\overline{G}(\vec x)$.
The weave is such that
$
\sum_{v\in V(\gamma)} \frac{8}{E(v)} \langle W,\xi|\hat{F}(v)\hat{G}(v)|W,\xi\rangle
= \sum_{{\rm Box}(\vec x)} F(\vec x) \sum_{v\in {\rm Box}(\vec x)}
\ \ell_P^3 \ \frac{8}{E(v)}\langle W,\xi|\frac{1}{\ell_P^3}\hat{G}(v)|W,\xi\rangle =
\int_{\Sigma} d^3x F(\vec x) \overline{G}(\vec x)$. 
Notice that the tensorial and Lie-algebra structure  should come out from flat spacetime quantities exclusively, i.e.  ${{^0}\!E}^{ia},\tau^k,\partial_b,\epsilon^{cde},\epsilon^{klm}$, where
$\delta^{ab}={{^0}\!E}^{ia}{{^0}\!E}_i{}^b$.  

In order to regain the flat spacetime kinetic term of the fermion Hamiltonian, we demand $|W,\xi\rangle$ to fulfill
\begin{equation}
\langle W,\xi|\xi_B(v)\frac{\partial\;}{\partial\xi^A(v)}
(\tau_k {\cal D}^{(\xi)}_c)^{AB} \hat{w}_{ia}(v) \hat{w}_{jb}(v) 
|W,\xi\rangle   
\approx \left(\frac{i}{\hbar}\xi_B(v)\pi_A(v)\ell_P {\cal L}^2\right)
\left(\frac{\ell_P^3}{{\cal L}^2} (\tau_k {\partial}^{(\xi)}_c)^{AB} \ {{^0}\!E}_{ia}(v) \ {{^0}\!E}_{jb}(v)\right)\;. \label{flatdel}
\end{equation}
The second parenthesis here dictates the overall structure: (\ref{DEFW}) indicates that each $\hat{w}_{ia}(v)$ contributes a factor $\frac{{\ell_P}^{3/2}}{\cal L}$, since the connection scales with $1/\cal L$
(large length limit $\Rightarrow$ flat spacetime), and $\sqrt{V}$ contributes a factor ${\ell_P}^{3/2}$. Independence on $\cal L$
of the final form of (\ref{flatdel}) gives the structure of the first parenthesis.
The notation $^{(\xi)}$ stands for acting only upon $\xi$. 
By expanding (\ref{EXPW}) at different orders in powers of $s$ and using (\ref{flatdel}), one can systematically determine all possible contributions. Some examples of correction terms are
\begin{eqnarray}
&& \frac{i}{4}\frac{{{\cal L}^2}}{\ell_P^3} \sum_{v\in V(\gamma)}\frac{8}{E(v)}
\epsilon^{ijk}\epsilon^{IJK} \ s_I{}^a s_I{}^b s_J{}^c s_K{}^d
\pi_A(v)  \partial_d \  \xi_B(v) \langle W,\xi|(\tau_k)^{AB} \left\{ \hat{w}_{iab}, \hat{w}_{jc} \right\}|W,\xi\rangle \nonumber\\
&\rightarrow&
\bar{\kappa}_5 \frac{\ell_P}{\cal L}
\int d^3x \frac{i}{2}\pi(\vec x)\tau_k  \ {{^0}\!E}^{kd}
\partial_d\xi(\vec x) \label{oscillation} \\
&& \frac{i}{4}\frac{{{\cal L}^2}}{\ell_P^3}   \sum_{v\in V(\gamma)}\frac{8}{E(v)}
\epsilon^{ijk}\epsilon^{IJK} \ \frac{1}{3 !}s_K{}^{a} \ s_K{}^{b} \ s_K{}^{c} s_I{}^d s_J{}^e
 \pi(v) \tau_k \ \partial_{a} 
\partial_b \partial_c \ \xi(v) \  \langle W,\xi |\hat{w}_{id}(v) \hat{w}_{je}(v)|W,\xi \rangle \nonumber\\
&\rightarrow&
- i \bar{\kappa}_8 \ \ell_P^2 \int d^3 x \ \pi(\vec x) \tau_k \ {{^0}\!E}{}^{k c} \  \partial_c  \ \nabla^2 \ \xi(\vec x) \;.
\label{helicities}
\end{eqnarray}

A similar treatment can be performed for every contribution to the Einstein-Dirac Hamiltonian constraint \cite{AMU}.
It is important to stress that
the prediction of the values of the corresponding coefficients $\kappa_i$
would require a precise definition of the flat weave (or even better  a Friedman-Lemaitre-Robertson-Walker weave), together with a detailed calculation of the matrix elements. Instead, within the present approach, the  neutrino equation, up to order $\ell_P^2$, becomes
\begin{eqnarray}
&&\left[i \hbar \frac{\partial}{\partial t}-i \ \hbar\ {\hat A }\ {\vec \sigma} \cdot \nabla +\frac{{\hat C}}{2\cal L} \right]\xi(t,{\vec x})
+m \left( \alpha -\beta \ i\hbar \ {\vec \sigma}\cdot \nabla \right)
 i \ \sigma_2 \  \xi^*(t,{\vec x}) = 0, 
\label{nueq}\nonumber \\
{\hat A}&=&\left(1 + {\kappa}_{1} \frac{\ell_P}{{\cal L}} 
+ {\kappa}_{2} \left(\frac{\ell_P}{{\cal L}} \right)^2+
\frac{{ \kappa}_3}{2} \ \ell_P^2 \ \ \nabla{}^2 \right), \quad  \alpha= \left(1 + { \kappa}_{8} \frac{\ell_P}{{\cal L}}\right), \label{effnueq}\\
{\hat C}&=&\hbar \ \left({ \kappa}_4 
+ { \kappa}_{5} \frac{\ell_P}{{\cal L}} 
+ { \kappa}_{6}\left(\frac{\ell_P}{{\cal L}} \right)^2
+\frac{{\kappa}_{7}}{2} \ \ell_P^2 \ \ 
\nabla{}^2 \right),\quad  \beta=\frac{ {\kappa}_9}{2\hbar} \ell_P. \nonumber
\end{eqnarray}
Notice that ${\kappa}_{4}$ would produce an additional Dirac mass for the neutrino. Since we are considering particles with a Majorana mass $m$, we take ${\kappa}_{4}=0$. In contrast to \cite{GP}, we have found no additional parity violation arising from the structure of the weave. 
The dispersion relation corresponding to (\ref{nueq}) is
\begin{equation}
\label{DR}
E_{\pm}^2(p, {\cal L})=(A^2+ m^2 \beta^2 )\ p^2 +m^2 \ \alpha^2+\left(\frac{C}{2{\cal L}}\right)^2 \ \pm \ B \ p,
\qquad { B}={ A}\left(\frac{{ C}}{\cal L} + 2 \alpha  \beta m^2 \right),
\end{equation}
where $A,B,C$ have been expressed in momentum space and depend on ${\cal L}$.
The $\pm$ in Eq. (\ref{DR}) stand for the two neutrino helicities. Let us emphasize that the solution $\xi(t,{\vec x})$ to Eq.(\ref{nueq}) is given by an appropriate linear combination of plane waves and helicity eigenstates, given that the neutrinos considered are massive.

Typically, for neutrinos, $\lambda_D <\!\!<\lambda_C$ and our approximation is meaningful only if ${\cal L} \leq \lambda_D$. In this way we make sure that Eq.(\ref{nueq}) is defined in a continous flat spacetime. From here 
on: $\hbar=c=1$. To estimate the corrections let us consider a massive neutrino with momentum ${\vec p}= {\vec {\bar p}}$. A lower bound for them is obtained by taking $1/ {\cal L}\approx 1/ \lambda_D =|{\vec {\bar p}}|={\bar p}$. Up to  leading order in  $\ell_P^2$,  we get
\begin{equation}
\label{DISP}
E_\pm({\bar p}):= E_\pm(p,{\cal L})|_{p={\bar p}, \ {\cal L}=1/{\bar p}} \approx {\bar p}+ \frac{m^2}{2{\bar p}} + \ell_P  \left( 
(\theta_2 \pm \theta_4) {\bar p}^2 +(\theta_1 \pm \theta_3)m^2 \right)  + (\theta_5 \pm \theta_6) \ell_P^2\ {\bar p}^3,
\end{equation}
where we are assuming that all $\theta_i$ are numerical quantities of order one. Besides, these are known functions of the original parameters $\kappa_i$ \cite{AMU}.
To  leading order in ${\bar p}$, the velocities are
\begin{equation} 
v_{\pm}({\bar p})=\frac{\partial E_{\pm}(p, {\cal L})}{\partial { p}}|_{p={\bar p}, \ {\cal L}=1/{\bar p}}= 1- \frac{m^2}{2 {\bar p}^2} + \kappa_1 (\ell_P  \ {\bar p}) \mp \kappa_7  \frac{(\ell_P \ {\bar  p})^2}{2} .
\end{equation}
The order of magnitude of the corrections arising from the present analysis is calculated using  the following values: $m = 10^{-9} \ {\rm GeV} $, $ {\bar p}\sim 10^5 \ {\rm GeV}$, $ L= 10^{10} \ {\rm l.y.}= 0.5\times 10^{42} \  \frac{1}{{\rm GeV}} $ (distance travelled by the neutrino from emission to detection on Earth). Let us observe that in Eq.(\ref{DISP}), the ratio between second and first order contributions in $\ell_P$  behaves like $({\bar p} \ell_P)\approx 10^{-14}$. Now consider the gravitationally induced time delay of neutrinos travelling at velocities $v_\pm$ with respect to those traveling at speed of light: $ {\Delta t}_{\nu}=|{L}(1-v_\pm)|= |\kappa_1| {L} \
({\bar p} \ \ell_P)$. Notice that this expression, though helicity independent, is of the same form as the one in Ref.\cite{GP} for photons. In our case we obtain   ${\Delta t}_{\nu}=0.3 \ |\kappa_1 |\times 10^{4}$ s. Besides, this correction dominates over the delay due to the mass term
$\frac{m^2}{2 {\bar p}^2}$ which is $\approx 10^{-10}$s. The second interesting parameter is the time delay of arrival for two neutrinos having different helicities: $\Delta t_{\pm}= L|(v_+ - v_-)|= |\kappa_7|
L \ ({\bar p}\ \ell_P)^2\approx  1.5 |\kappa_7|\times 10^{-11}$ s. This correction is supressed by a factor of $({\bar p}\ \ell_P) $ with respect to the former and it is comparable to the time delay caused by the mass term. Finally, consider the  characteristic length $L_{\rm os}$ corresponding  to  two-flavour neutrino oscillations, given by
$L_{\rm os}= \frac{2 \pi }{(E_a-E_b)}\equiv \frac{2 \pi }{\Delta E}$, where $E_{a,b}$ denotes de energy corresponding to the mass eigenstates
of  the neutrinos  with masses $m_{a,b}$ respectively. As usual, we assume that neutrinos are highly relativistic (${\bar p}_a >\!\!> m_a $) and also  that ${\bar p}_a \sim {\bar p}_b ={\bar p }\sim E$. The phase $\Phi_{\rm os }$ describing the oscillation is $\Phi_{\rm os }= \frac{\pi L}{L_{\rm os}}$, where $L$ is the distance travelled by the neutrino between emission and detection. The energy difference for the corresponding two flavours is 
\begin{eqnarray}
\label{DELTAE}
\Delta E= \frac{\Delta m^2}{2 {\bar p}} + \Delta \rho_1 {\bar p}^2 \ell_P +
\Delta (\rho_2 m^2) \ell_P
\approx  \left( 10^{-26} + \Delta \rho_1\times 10^{-9} +  10^{-40} \right)\ {\rm GeV}.
\end{eqnarray}
This result could yield bounds upon ${\Delta \rho}_1$, which measures
a violation of universality in the  gravitational coupling for different neutrino flavours. For the above estimation, $\Delta m^2 \approx 10^{-21} \ {\rm GeV^2}$, and $\Delta (\rho_2 m^2) \approx
\Delta m^2$ were used. Here $\rho_i$ are flavour dependent quantities of order one. To conclude, we notice that ($\ref{DELTAE}$) implies
$L_{\rm os}^{-1} \propto {\bar p}^2 \ell_P $, seemingly an effect not
considered previously \cite{ITA}. 

{\bf Acknowledgments}. We thank R. Gambini for useful discussions 
about \cite{GP} and T. Thiemann for illuminating comments
on the regularization problem.
Partial support is acknowledged from CONICyT-CONACyT  E120-2639,
DGAPA IN100397 and CONACYT 3141PE.
The work of JA  is partially supported by Fondecyt
1980806
and the CONACyT-CONICyT project 1997-02-038. We also acknowledge
the
project Fondecyt 7980018.


\end{document}